\documentclass[twocolumn]{aastex631}
\hypersetup{linkcolor=red,citecolor=blue,filecolor=cyan,urlcolor=magenta}
\shorttitle{Sherpa Python Package}

\newcommand{\sherpa}{{\sl Sherpa}}
\newcommand{\chandra}{{\sl Chandra}}

\begin{document}

\title{Sherpa: An Open Source Python Fitting Package}

\correspondingauthor{Aneta Siemiginowska}
\email{asiemiginowska@cfa.harvard.edu}

\author[0000-0002-0905-7375]{Aneta Siemiginowska}
\affiliation{Center for Astrophysics $|$ Harvard \& Smithsonian, 
60 Garden St., Cambridge, MA, 02138}

\author[0000-0003-4428-7835]{Douglas Burke}
\affiliation{Center for Astrophysics $|$ Harvard \& Smithsonian, 
60 Garden St., Cambridge, MA, 02138}

\author[0000-0003-4243-2840]{Hans Moritz G\"unther}
\affiliation{MIT Kavli Institute for Astrophysics and Space Research, Massachusetts Institute of Technology, Cambridge, Massachusetts, United State}

\author[0009-0006-5274-6439]{Nicholas P. Lee}
\affiliation{Center for Astrophysics $|$ Harvard \& Smithsonian, 
60 Garden St., Cambridge, MA, 02138}

\author{Warren McLaughlin}
\affiliation{Center for Astrophysics $|$ Harvard \& Smithsonian, 
60 Garden St., Cambridge, MA, 02138}

\author[0000-0002-7939-377X]{David A. Principe}
\affiliation{MIT Kavli Institute for Astrophysics and Space Research, Massachusetts Institute of Technology, Cambridge, Massachusetts, United State}

\author{Harlan Cheer}
\affiliation{Center for Astrophysics $|$ Harvard \& Smithsonian, 
60 Garden St., Cambridge, MA, 02138}

\author[0000-0002-6414-3954]{Antonella Fruscione}
\affiliation{Center for Astrophysics $|$ Harvard \& Smithsonian, 
60 Garden St., Cambridge, MA, 02138}

\author{Omar Laurino}
\affiliation{Center for Astrophysics $|$ Harvard \& Smithsonian, 
60 Garden St., Cambridge, MA, 02138}

\author[0000-0002-7093-295X]{Jonathan McDowell}
\affiliation{Center for Astrophysics $|$ Harvard \& Smithsonian, 
60 Garden St., Cambridge, MA, 02138}

\author{Marie Terrell}
\affiliation{Center for Astrophysics $|$ Harvard \& Smithsonian, 
60 Garden St., Cambridge, MA, 02138}





\begin{abstract}

We present an overview of \sherpa, an open source Python project, and discuss its development history, broad design concepts and capabilities.
\sherpa\ contains powerful tools for combining parametric models into complex expressions that can be fit to data using a variety of statistics and optimization methods. It is easily extensible to include user-defined models, statistics, and optimization methods. It provides a high-level User Interface for interactive data-analysis, 
such as within a Jupyter notebook, and it can also be used as a library component, providing fitting and modeling capabilities to an application. We include a few examples of \sherpa\ applications to multiwavelength astronomical data.

\end{abstract}




\section{Introduction} \label{sec:intro}

Data processing software developed by astronomical observatories typically generates standard data products which constitute the basis for scientific analysis.
Example data products include spectra and images, data cubes, or for the case of high-energy astrophysics, lists of detected photons called \emph{event lists}. Generally, these software packages focus on raw data reduction leaving scientific inference - which requires appropriate statistical methodology and theoretical understanding of the astrophysics - to other specialized packages. While the underlying astrophysics theory must be specific to the nature of the observed source, the statistical methodology is broader. This methodology involves applying the theory to the observations and making scientific inference based of the quality and characteristics of the data. 

\sherpa\, is a general modeling and fitting application with a library of models, statistics and optimizers. \citep{Freeman2001, Refsdal2009}. It was developed by the \chandra{} X-ray Center (CXC) with a focus on fitting X-ray spectra and images. 
It is distributed as part of the CIAO (\chandra\ Interactive Analysis of Observations) software package \citep{Fruscione2006}
and as a stand-alone Python package under the open-source GNU General Public License \citep{zenodo}. 
While \sherpa{} was originally developed with the specific requirements of X-ray data modeling in the Poisson regime, which use calibration information in forward fitting methods, it is designed to handle more generic data such as optical spectra or spectral energy distributions (SEDs) \cite[e.g.,][]{AGNPY2022, Fantasy2023}. In the case of 2-D image analysis, \sherpa{} can incorporate ancillary data from background maps, exposure maps and point spread function (PSF) images as part of 2-D model expressions. Accounting for these factors is necessary for robust scientific inference.
\sherpa{} has been used for analysis of multiwavelength images, spectra, and time series from many telescopes, including ground-based optical telescopes \cite[e.g.,][]{SDSS2022, OpticalSpectra2024, C2Lines2024}, 
the Hubble Space Telescope (HST)\citep{HSTCOS2023}, the \emph{Spitzer}-IRS infrared telescope \citep{Spitzer2013}, \chandra\, XMM-{\it Newton} and NuStar in X-rays, the Fermi Space Telescope for high energy $\gamma$-rays \citep{GammaTeV2024} and H.E.S.S. for TeV data \citep{HESS2012}. It can also be used easily with non-astronomical data \cite[e.g.,][]{Aldcroft2010}. \sherpa{} is flexible, modular and extensible. It has an IPython user interface and it is also an importable Python module. \sherpa{} models, optimization and statistic functions are available via both C++ and Python for software developers using such functions directly in their own code (e.g., the recent \texttt{fantasy} tool for fitting optical spectra by \citealt{Fantasy2023}).

Many of \sherpa's users are in the X-ray community, but its Python foundations and compatibility with multi-wavelength data makes it an attractive and viable option for astronomers working at other wavelengths. 
\sherpa\ handles WCS (World Coordinate System) information and calibration products, has the fit statistics appropriate for both Poisson and Gaussian data, and contains extensively tested robust optimization methods. 

In the following sections we present a brief history of the \sherpa\ development (Sec.~\ref{sec:history}), a quick study of it's use by citation numbers (Sec.~\ref{sect:pub}), an overview of the design (Sec.~\ref{sec:design}), its current capabilities with 
examples (Sec.~\ref{sec:capabilities}), \sherpa's place in the Python ecosystem (Sec.~\ref{sec:ecosystem}), and the open source management of the project (Sec.~\ref{sec:manage}). 
This paper provides an overview of the project and it is not intended to be software documentation. The open source code and details of the releases are available on GitHub\footnote{\url{https://github.com/sherpa/sherpa}}. The user documentation, reference/API with class diagrams are located on the ReadTheDocs\footnote{\url{https://sherpa.readthedocs.io/en/latest/}} site.

\section{A brief history of \sherpa{}}
\label{sec:history}

\sherpa{} has been developed over the past two decades by the \chandra{} X-ray Center (CXC) \citep{Doe1997,Doe1998,Freeman2001}
as a complement to the 
\chandra\ Interactive Analysis of Observations (CIAO)
software \citep{Fruscione2006}. The main goal was to provide \chandra\ users with a general fitting application with flexibility in the selection of statistics and optimization methods and capabilities for building complex model expressions. Although, the initial focus was on fitting X-ray data,
the intent was to develop a general application and promote a multi-wavelength analysis of diverse datasets.

\sherpa{} was originally written in a combination of C, C++ and Fortran. To adapt the user interface to the evolving needs of the astronomical community, a major re-write of the \sherpa\ software was undertaken \citep{Doe2006}. This rewrite, initially released as a Beta version in 2007, re-implemented \sherpa\ in Python with some modules in C++ for improved speed.\citep{Doe2007,Refsdal2009,Laurino2015,Laurino2019}. 
Since 2014, \sherpa{} follows an open-development model with the source code and all development occurring on GitHub. It has had a total of 23 public open source software releases to date (\url{https://github.com/sherpa/sherpa})\citep{zenodo}. It is open source code licensed under the GNU General Public license (GPL) version 3 . \sherpa\ can be downloaded as part of the CIAO software distribution or as a Python package from GitHub or Pypi.

\begin{figure}[b!]
\centering
\vspace{3pt}
  \includegraphics[width=0.42\textwidth]{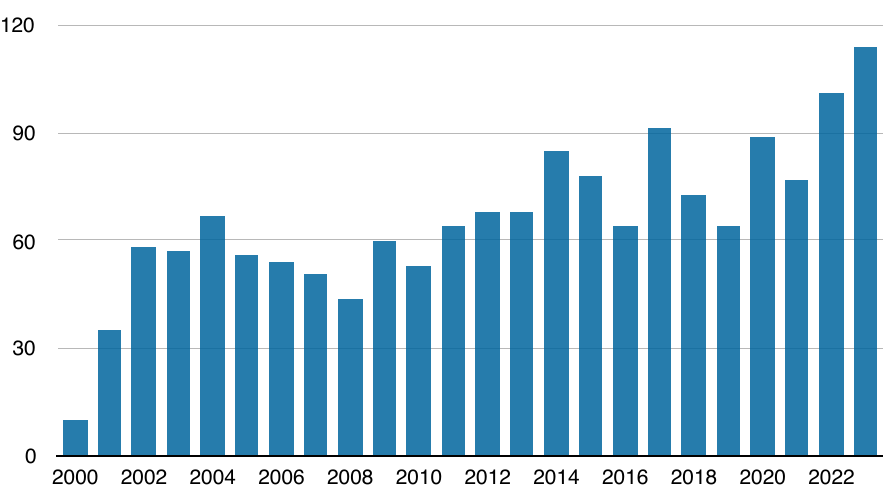} \hspace{10pt}
  \includegraphics[width=0.44\textwidth]{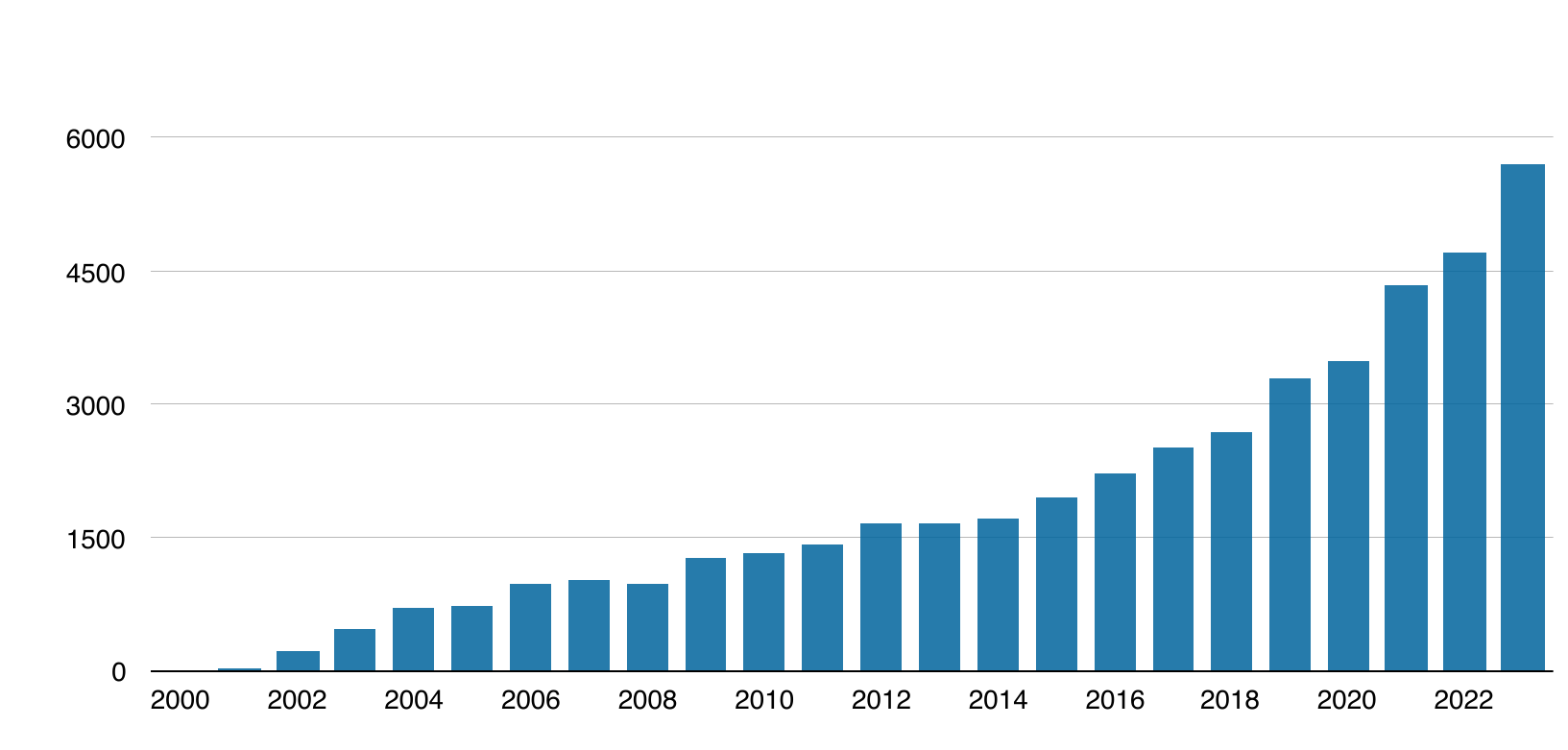}
  \vspace{5pt}
  \caption{Publications that used \sherpa{} (top panel) and the citations of these papers (bottom panel) over time till the end of 2023 (from the ADS library). The x-axis shows years and the vertical axis shows the number of publications.
   }
  \label{fig:papers}
\end{figure}

\section{Publications}
\label{sect:pub}
A fulltext search in the Astrophysics Data System (ADS\footnote{\url{https://ui.adsabs.harvard.edu/}}) returns more than 1,600 references to \sherpa\ being used in scientific publications.
Figure~\ref{fig:papers} (top panel) presents yearly science publications (until Dec. 2023) that used \sherpa\ and indicates that the program's usage has been steady with an increasing number of publications in the past 2-3 years. The bottom panel in Figure~\ref{fig:papers} shows the increasing citations of these papers with a total of 36,417 cited papers.

Areas of research covered by the scientific publications are diverse. Some recent examples of \sherpa\ applications include: 
\\
(1) analysis of emission lines in optical spectra of an AGN sample
    \citep{OpticalSpectra2024};\\
(2) modeling the high resolution optical spectral lines from molecular clouds and synthetic spectra, in the EDIBLES Very Large Telescope survey \citep{C2Lines2024}; \\
(3) fitting complex models to X-ray spectra of AGN, SNR or $\gamma$-ray source classifications \citep{AGN2024,Tycho2024,SpectraBXA2024} together with the Bayesian X-ray Analysis package \citep[BAX,][]{BXA}; 
(4) modeling clusters of galaxies using optical, radio and X-ray images of gravitationally lensed clusters of galaxies \citep{Lens2024,Ebeling2024};\\
(5) modeling $\gamma$-rays and TeV spectra with the \texttt{gammapy} software using \sherpa's \texttt{simplex} optimization method \citep{GammaTeV2024}.\\

\chandra\ observers use \sherpa\ to fit low- and high spectral resolution X-ray spectra, images, and radial profiles. 
For advanced analysis tasks \sherpa\ can be and has been incorporated into processing scripts and pipelines.
At the CXC \sherpa\ was used in data processing to fit $>$1.3 million individual source detections and performed approximately a million additional image fitting operations to compile the \chandra\ Source Catalog\footnote{\url{https://cxc.cfa.harvard.edu/csc/}} \citep{Evans2010,Evans2024}.
\sherpa\ was embedded into Iris, a Virtual Observatory
application for modeling spectral energy distributions \citep{Iris2014}.
Outside the CXC, 
\cite{Lens2024} recently incorporated \sherpa\ into a pipeline for modeling a mass of a gravitational lens with X-rays, lens parameters and galaxy kinematics.  
In this case, \sherpa\ was used to fit a photoionization model to obtain a temperature map of an X-ray cluster. In another example, \cite{Cavity2024} fit a 2-D X-ray image of a galaxy with a smooth 2-D $\beta$-model and used the 2-D-image residuals to detect 
cavities to produce a training data set for the machine learning application CaDeT.

\begin{figure}
\centering
  \includegraphics[width=0.9\columnwidth]{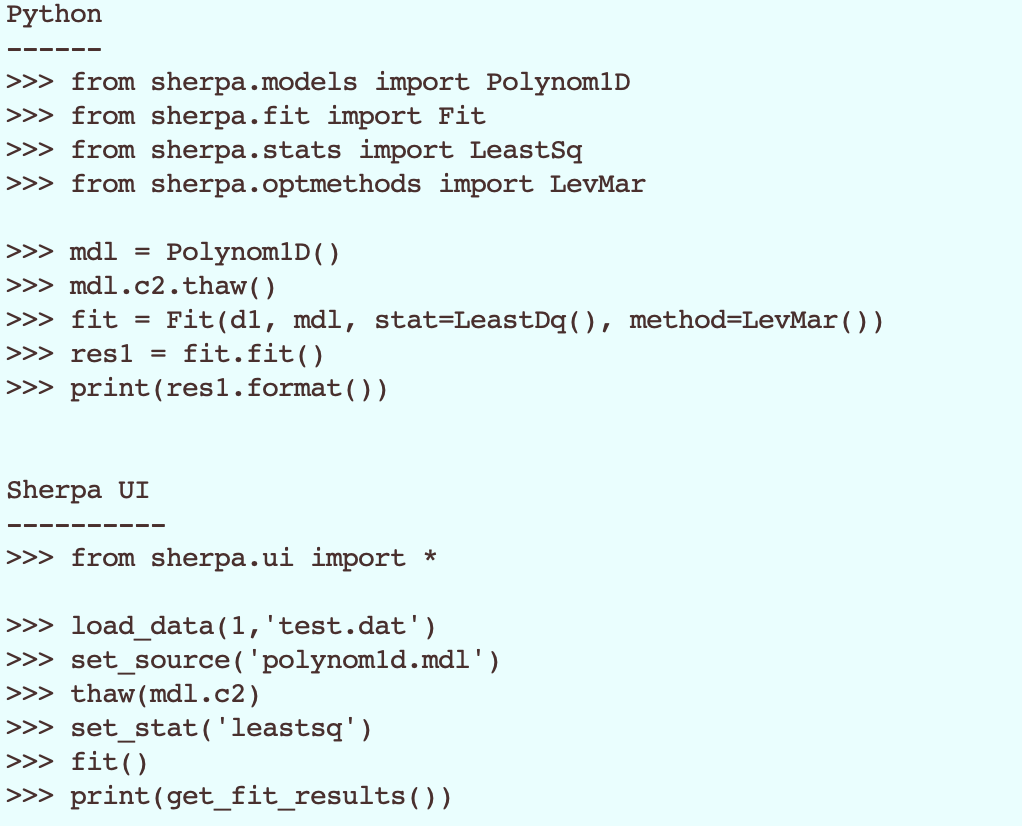}
  \vspace{5pt}
\caption{An example of \sherpa\ Python commands for fitting a polynomial to 1D data array, \texttt{d1}, accessing Python objects (top) or using the UI (bottom)}
\vspace{5pt}
\label{fig:simple}
\end{figure}

\section{Design and Documentation}
\label{sec:design}

\sherpa's design has two main components: a user interface (UI) session and a Python object layer \citep{Doe2007}. This provides flexibility and allows \sherpa\ to be used in Python scripts, Python and IPython shells, Jupyter notebooks and in other applications, such as e.g. \texttt{ds9 DAX} \citep{DAX}.

The \sherpa\ user interface (UI) was designed more than two decades ago to simplify user interactions on the command line, within a \sherpa\ interactive session. It was based on the standard 
style used by IRAF \citep{IRAF} and XSPEC \citep{XSPEC} at that time. As a result, there are many command line functions that can be used directly within the session or included in command-line scripts. This classical user interface allows CIAO users to open a standard \sherpa\ session in the IPython environment and have user-friendly access to the defined functions for accessing data, selecting models, statistics, and optimization methods. This interface also provides many standard visualization options making it easy to generate and display standard plots and images. The session default settings can be customized to access additional options for I/O, plotting
statistics etc.

The Python object layer is designed for power users who want to access Python modules directly, write complex Python scripts which include \sherpa, or use parts of \sherpa\ in their own Python package.

This bi-modal design also impacted the design of the user documentation. Originally, the \sherpa\ documentation was developed within the CIAO environment. In this system, XML files contain descriptions of UI functions and form a base for online help (\texttt{ahelp}) in the CIAO-\sherpa\ session. The CIAO-\sherpa\ website documentation is based on the same XML files.
and provides an extended description of concepts and examples in form of \textit{analysis threads}. The CIAO-\sherpa\ website has been available to users since the early days of the \chandra\ mission. Although it focuses on specific X-ray analysis issues,  it also contains  information and directions for modeling and fitting astronomical data in general.

Documentation of the \sherpa\ Python package is built directly from the content included in the \sherpa\ code and documentation files in the \texttt{docs} directory on GitHub. 
The standard Python \texttt{sphinx} system is used to generate \sherpa\ \texttt{ReadTheDocs} pages.
The content was initially derived from the CIAO-\sherpa\ documentation and it evolved to expand content at the code level and improve code accessibility.
More recently, several Jupyter Notebooks have been added to provide interactive examples.

\section{Capabilities}
\label{sec:capabilities}

\begin{figure}
\centering
  \includegraphics[width=0.5\textwidth]{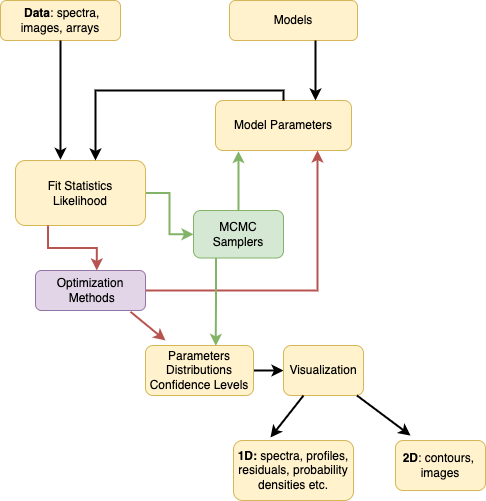}
  \vspace{5pt}
\caption{\sherpa\ forward fitting. 1-D or 2-D data and complex models are defined. The best fit parameters are found by a classic optimization method (violet box and red connectors) or model parameters are sampled from the posterior distribution with a Bayesian MCMC sampler (green box and connectors). The parameters with the confidence levels and the distributions can be visualized with the plotting and imaging options.}
\vspace{5pt}
\label{fig:design}
\end{figure}

Figure~\ref{fig:design} displays the main \sherpa\ components with options for data, models, optimization methods, Bayesian samplers and returned parameter distributions which can be visualized. The details are available on the \sherpa\ ReadTheDocs
documentation website.

The following sections describes the main capabilities of \sherpa.

\subsection{Data}

\sherpa\ can create, store, filter, and display data. It distinguishes between data that is binned (such that the value y represents an observations or model in an interval of x values ranging from xlo to xhi) and data where y is the value at point x.
There are three main data classes defined in the \texttt{sherpa.data} module:
Data1D, Data1DInt, and Data2D to handle respectively (x, y), (xlo, xhi, y), and (x1, x2, y) data. Dynamic filtering and grouping allows users to select a subset of the data range, or group the data for model evaluation on different grids.

The \texttt{sherpa.astro.io} module contains routines to read and write FITS\footnote{{\url{https://heasarc.gsfc.nasa.gov/docs/heasarc/fits.html}}} (Flexible Image Transport System)
and ASCII format data. \sherpa\ supports a choice of two I/O backend packages: \texttt{Astropy} \citep{astropy,astropy_II,astropy_III} and \texttt{crates} (in CIAO).
Standard astronomy data files, such as spectra, images, and calibration files (e.g. ARF or RMF in the X-ray regime; see OGIP/92-007\footnote{\url{https://heasarc.gsfc.nasa.gov/docs/heasarc/ofwg/docs/spectra/ogip_92_007/ogip_92_007.html}} and CAL/GEN/92-002\footnote{\url {https://heasarc.gsfc.nasa.gov/docs/heasarc/caldb/docs/memos/cal_gen_92_002/cal_gen_92_002.html}} for file format definitions) are read or written using the selected backend package.

\subsubsection{Astronomical Data}

The \texttt{sherpa.astro.data} module is tailored for astronomy datasets and supports two types of data: two-dimensional images (DataIMG) and X-ray spectra (DataPHA),
along with associated calibration information (DataARF and DataRMF).
Both types of data extends the module's capabilities to meet the specific requirements of astronomical data and support the following:

- image filtering using geometric shapes (regions);

- different coordinate units for filtering images (logical, physical, and WCS), depending on the available metadata;

- different analysis units  (channels, energy, and wavelengths) for filtering and displaying X-ray spectral files in the PHA format 

- dynamical re-binning of PHA data to improve the signal to noise (grouping and quality).

\subsection{Building Complex Models}

The \texttt{sherpa.models.model} module allows models to be defined and combined 
using allowed expressions, such as addition, subtraction, and multiplication. The model parameters can be linked across different models and datasets. 

The \texttt{sherpa.models} and \texttt{sherpa.astro.models} modules contain a library of 1-D and 2-D models. Additionally, the specialized models are provided in \texttt{sherpa.astro.optical}, \texttt{sherpa.astro.xspec}, \texttt{sherpa.instrument}, and \texttt{sherpa.astro.instrument}.

The instrument modules contain definitions of specific calibration models, which are typically input from files such as ARF and RMF for X-ray spectral fitting. By default, these calibration models are convolved with the physical models during the forward fitting of the data.

All models specific to X-ray spectra and contained in the 
\texttt{sherpa.astro.xspec} module are provided by the XSPEC model library \citep{XSPEC}, which can be linked to \sherpa\ at build time. 

\subsection{Simulations}
\label{sec:sims}

\sherpa\ provides several methods to generate simulated datasets using the \texttt{numpy.random} library.
The \texttt{sherpa.ui.fake} function generates the synthetic 1D or 2D data for a model, assuming a default Poisson distribution. The \texttt{sherpa.astro.ui.fake\_pha} function generates X-ray spectra for a given spectral model, exposure time and the instrument response files.

The \texttt{sherpa.sim} module provides the Bayesian MCMC algorithm for low counts Poisson data. It also includes the \texttt{sherpa.sim.simulate} module with classes to support simulations for the posterior predictive p-values and likelihood ratio tests for spectral models \citep[see][]{Protassov2002}

\subsection{Fitting}
\label{sec:fitting}

The \texttt{sherpa.fit} module represents the core of \sherpa's fitting functionality. The \texttt{Fit} class combines data and model expression to be fit and uses an optimizer to minimize the selected statistics function.

\sherpa\ uses the forward fitting method to determine the best-fit parameters.
The statistical functions for modeling Poisson and Gaussian data are available 
in the \texttt{sherpa.stats} module.
Several robust optimizers specifically designed for fitting non-linear 
models are provided in the \texttt{sherpa.optmethods} module, including Levenberg--Marquardt \citep{Levenberg,Marquardt}, Nelder-Mead Simplex \citep{NelderMead1965} and Monte Carlo/Differential Evolution \citep{StornPrice1997}.
These optimizers work by optimizing the fit statistics function and returning the best-fit model.

Several methods are available in the \texttt{sherpa.estmethods} module for computing the uncertainties and confidence level for the best-fit parameters. The \texttt{confidence} method computes confidence bounds 
by varying a parameter along the grid and optimizing the model for the other parameters to find  the best-fit statistics, though it can be relatively slow due to the large number of fitting operations required.  
The \texttt{covar} method, on the other hand, is based on the covariance matrix and does not account for correlations between the parameters, but is much faster.

\subsection{Bayesian Analysis}
 
\sherpa's implementation of Bayesian analysis with Poisson Likelihood and priors is based on the model described in \cite{VanDyk2001} and implemented in the \texttt{pybloxcs} package \citep{Siemiginowska2011}. It uses the Metropolis or Metropolis-Hastings algorithm in the MCMC (Markov-Chain Monte Carlo) to sample the posterior density. The main function which runs the MCMC is defined in \texttt{sherpa.astro.ui.get\_draws}. Calibration uncertainties can be incorporated into the sampling by using the Pragmatic or Full Bayes approach, as described in \cite{Lee2011} and \cite{Xu2014} using calibration files, such as ARF and RMF in the X-ray case, which include uncertainties.

\subsection{Visualization}
\label{sec:visualization}

\sherpa\ has a generic plotting abstraction and is currently released with three interfaces: (1) \texttt{matplotlib} \citep{Hunter:2007} for publication-quality static plots, (2) \texttt{bokeh} for interactive plots, and (3) \texttt{ds9} for visualizing data using the interactive \texttt{ds9} application \citep{ds9}.

The \texttt{sherpa.plot} module defines the main plot classes, such as \texttt{plot}, \texttt{histogram}, \texttt{point}, \texttt{contour}, and \texttt{image}. It also contains many options for common display options for data, model fit and residuals, and other pre-defined options to support, for example, confidence contours or probability density histograms.

A separate \texttt{sherpa.astro.plot} module defines classes for plotting astronomical data and includes specific options for plotting PHA data or displaying FITS images.

\section{Fitting Examples}
In this section, we present some examples of \sherpa\ usage.

\subsection{Fitting 1D data}

\sherpa\ can fit 1D data (simultaneously or individually), binned or unbinned, including: spectra, surface brightness profiles, light curves and general ASCII arrays. The standard PHA type files for X-ray spectra and the associated calibration files are handled in a special manner, as the required information is automatically read from the files' headers. Multiple data sets can be fit simultaneously with model parameters varied independently or linked across data sets.
The \sherpa\ documentation provides many examples of analysis threads and usage to guide users.

\begin{figure}
    \centering
    \includegraphics[width=0.48\textwidth]{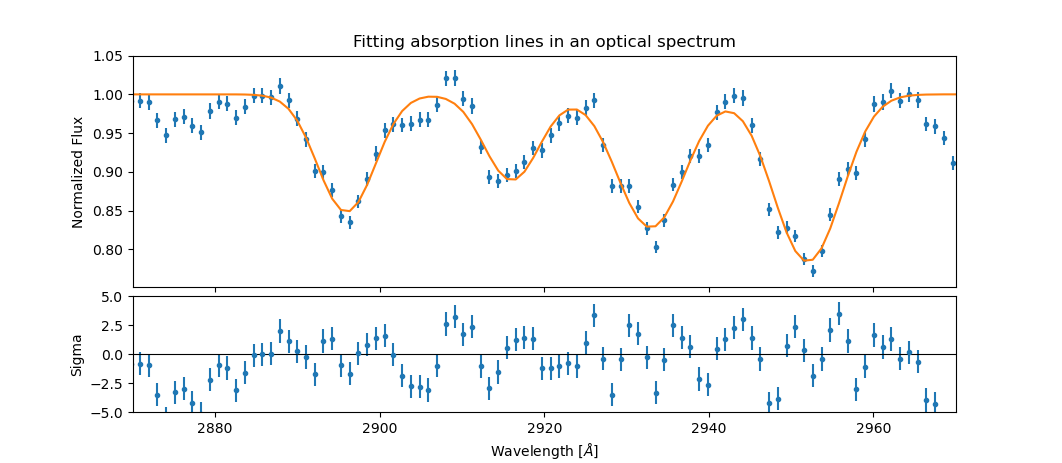}
    \caption{An example of fitting multiple absorption lines to the optical spectrum. The model consisted of a polynomial ({\texttt{polynom1d}}) and four independent absorption lines ({\texttt{opticalgaussian}} parameterized by the position, width and optical depth). The top panel shows the data points (blue) overplotted with the best fit model with the residuals shown in the bottom panel.}
    \label{fig:optical}
\end{figure}

\begin{figure}
    \centering
    \includegraphics[width=0.48\textwidth]{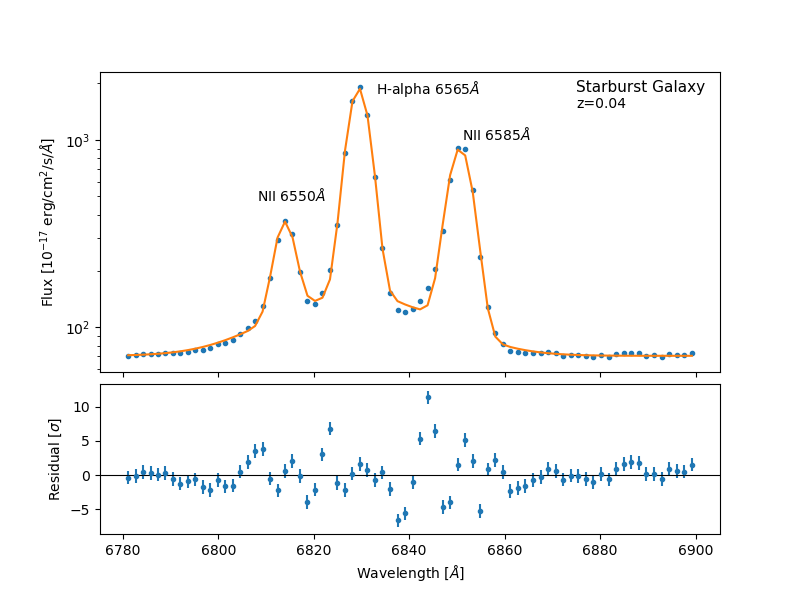}
    \caption{An optical spectrum (SDSS) fit with a model combined with a constant model ({\texttt{const1d}}) and a set of four Gaussian emission lines ({\texttt{gauss1d}}). The top panel shows the data (blue points) and the model fit (orange) with a sum of three Gaussian components overlayed with grey dashed line. The lower panel shows the residuals. }
    \label{fig:sdss}
\end{figure}

\begin{figure}
    \centering
    \hspace{-0.3in}
    \includegraphics[width=0.65\columnwidth]{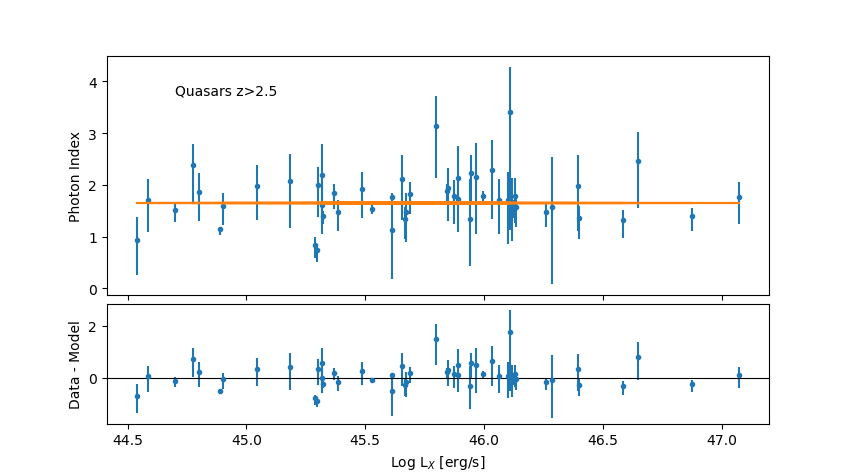}
    \hspace{-0.2in}
    \includegraphics[width=0.38\columnwidth, height=3cm]{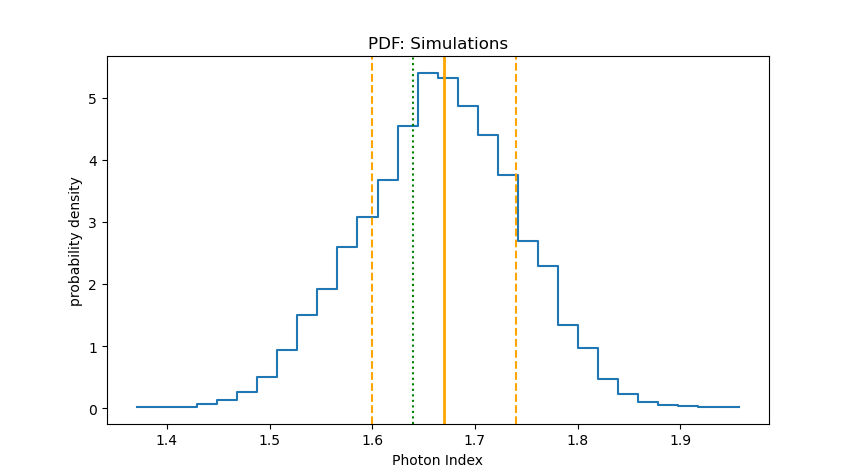}
    \caption{Fitting luminosity vs. photon index relation for a sample of quasars (\cite{Shaban2022}). Photon index data have asymmetric errors. {\bf Left:} The upper panel displays the photon index (blue points with asymmetric errors for each source at a given luminosity) and the best fit linear model (orange line).  The lower panel shows the residuals. {\bf Right:} Probability distribution from the bootstrap. The orange lines mark the median and 1$\sigma$ uncertainties. The green dotted line marks the best fit model assuming an average measurement error given by upper and lower error. }
    \label{fig:errors}
\end{figure}

\begin{figure*}
    \centering
    \includegraphics[width=0.49\textwidth]{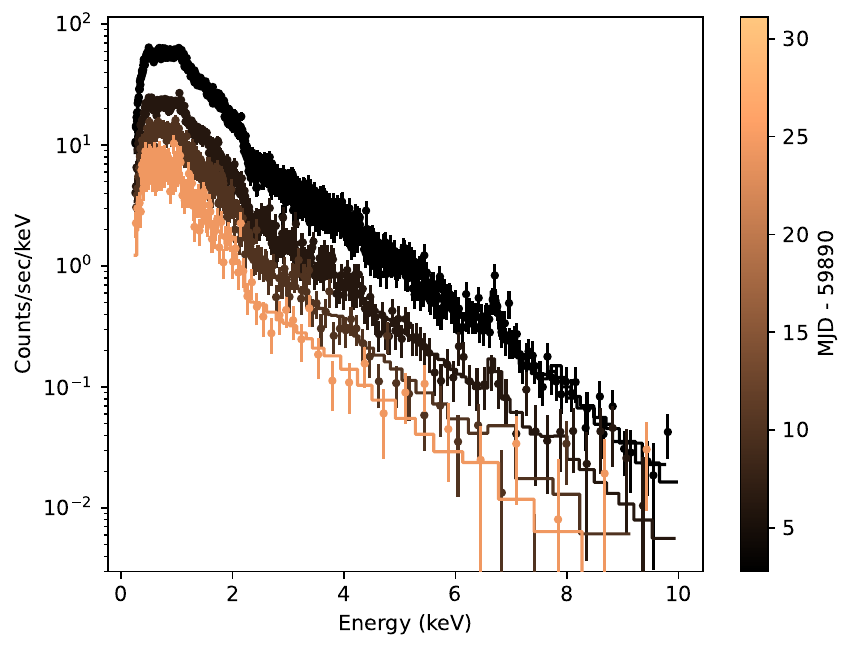}
    \includegraphics[width=0.49\textwidth]{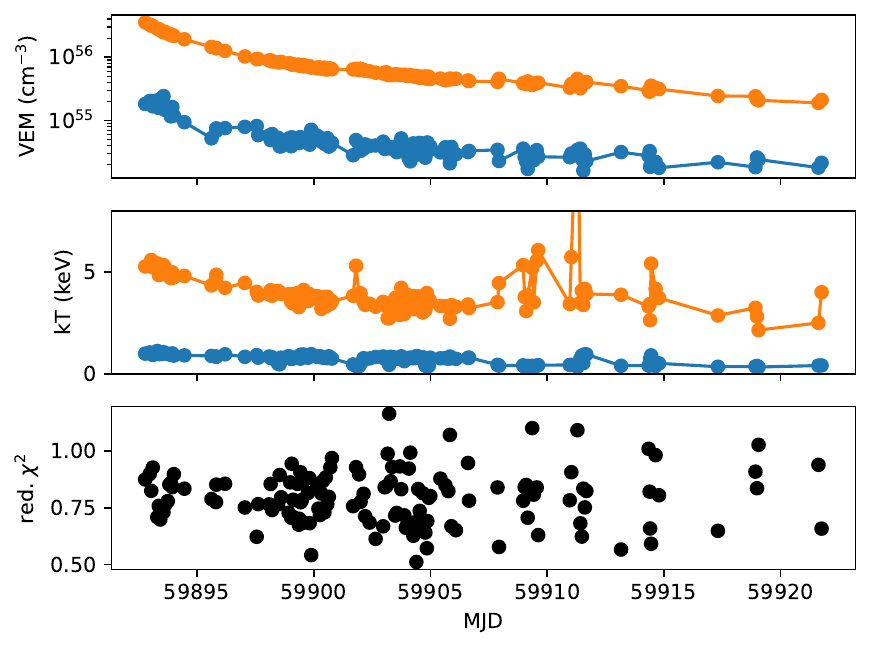}
    \caption{NICER observations of a large stellar flare from a K-type giant star with \sherpa\ fits. \textbf{Left:} Four selected examples from the set of 127 spectra. Each spectrum is fitted with the same type of model (two thermal emission components), but each fit is independent with four parameters (temperature and volume emission measure [VEM] for both components). Color bar shows the time of observation (see color bar on the left) and indicates the flare decay over time (black to orange). \textbf{Right:} Fitted VEM (top) and temperature k$T$ (center) over time (k the Boltzmann constant) for the two emission components (orange and blue). The fit quality (reduced $\chi^2$) for each individual fit is shown in the bottom panel. Error bars are not shown for clearity.}
    \label{fig:NICER}
\end{figure*}

Figure~\ref{fig:optical} illustrates a fit to an optical spectrum (a simple ASCII table).
A small section of the SDSS\footnote{Sloan Digital Sky Survey} spectrum of a starburst galaxy, SDSS\,J151806.13+424445.0\footnote{\url{https://skyserver.sdss.org/dr18/VisualTools/quickobj}} ($z=0.0403$) is fitted assuming a constant model and multiple Gaussian profiles for the emission lines in Figure~\ref{fig:sdss}. The resulting data and model spectrum displayed the redshifted emission lines at the observed wavelength position for NII\,6568$\AA$, H${\alpha}$\,6565$\AA$, NII\,6583$\AA$.

An example of a linear model fit to data with asymmetric errors is shown in Figure~\ref{fig:errors}. 
This is often the case when the data are provided as a result of the previous analysis. In this example high redshift quasars were fit assuming an absorbed power law model (\citep{Shaban2022}. The best fit photon index has asymmetric uncertainties, so the lower and upper bounds have different values.
In order to evaluate a dependence of the photon index on the X-ray luminosity for this quasars sample a constant model was fit to the data assuming measurement errors given by an average of upper and lower errors (shown as orange line in Figure~\ref{fig:errors}). However, a bootstrap procedure ({\texttt{resample\_data}} in \sherpa), can be used to simulated the data assuming a skewed distribution for the measurement errors (e.g. a sum of two Gaussians with different width set by the lower and upper errors).
The resulting median value of the photon index and the uncertainties obtained through bootstrap methods indicate a slight bias in fitting these data assuming an average errors as illustrated in the right panel in Figure~\ref{fig:errors}.

Figure~\ref{fig:NICER} shows a typical use of Sherpa fitting 1D data, in this case X-ray spectra observed with the the Neutron Star Interior Composition Explorer \citep[NICER,][]{2017NatAs...1..895G} from the flaring K giant HD~251108. \emph{NICER} started observing the source after a flare peak and observed for about 20~days as the flare cooled down. In a first step all 127 spectra are fit independently 
with the same model (a sum of two collisionally excited, optically thin plasma models from APEC \citep{2012ApJ...756..128F}).
At later times, when the flare is faint, the fit uncertainties are larger,
but 
\sherpa\ fits indicate that we observe the cooling of the flare plasma over the entire range shown in Fig.~\ref{fig:NICER}; see \citet{guenther_submitted} for details of this dataset.

\subsection{Fitting 2D data}

\sherpa\ can handle 2D images and includes more than ten spatial models. X-ray images are typically sparse and require Poisson likelihood which is one of the available choices of the statistical method. Imaging calibration data, such as point spread function (PSF), exposure maps, and background maps can be included in \sherpa's 2D forward fitting.

\begin{figure*}
    \centering
    \includegraphics[width=.9\textwidth]{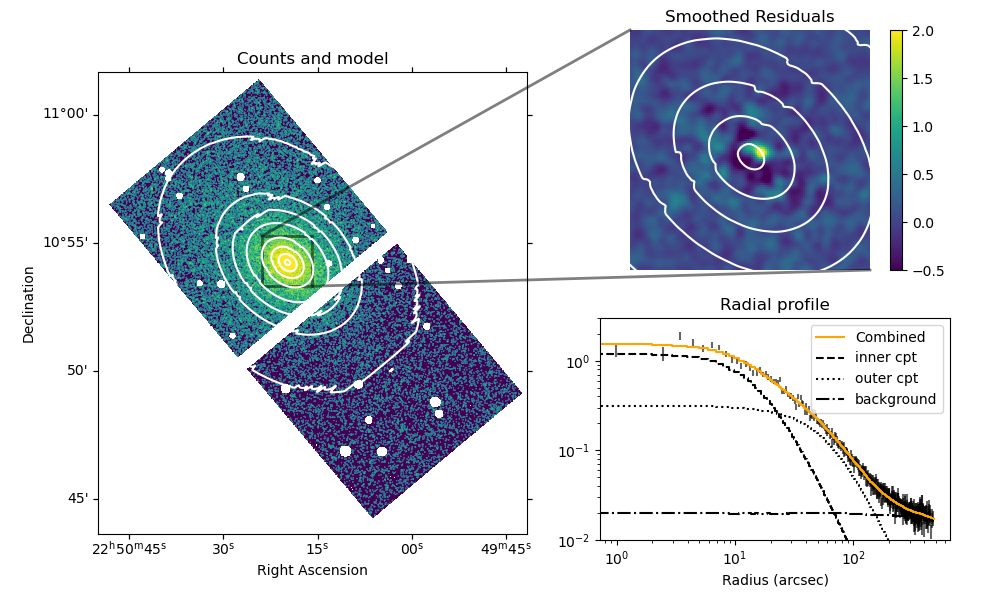}
    \caption{Main left image: \chandra\ image of an X-ray cluster A2495 overlayed with the contours of the 2D fit assuming two \texttt{beta2d} model components with the same location of the center. Top right panel shows the smoothed residuals with visible offset from the central location of the model. This cluster is known to have offsets between the central galaxy and the hot gas. The bottom panel shows the radial profiles of the surface brightness based on the fitted model components, the sum of the two component overlayed over the data and the background level.}
    \label{fig:example2}
\end{figure*}

Figure~\ref{fig:example2} shows an example of 2D fitting of the X-ray cluster A2495 observed with \chandra\ (\cite{Rosignoli2024}). The system is known for the triple offsets between the center of the intercluster medium (ISM), the central galaxy (BCG), and the warm gas. 
The 2D model for the surface brightness consisted of two \texttt{beta2d} model components with the same location of the center. The residual image in Fig.~\ref{fig:example2} shows an additional peak offset from the center of the model indicating a presence of offset emission.  

\begin{figure}
    \centering
    \includegraphics[width=0.23\textwidth]{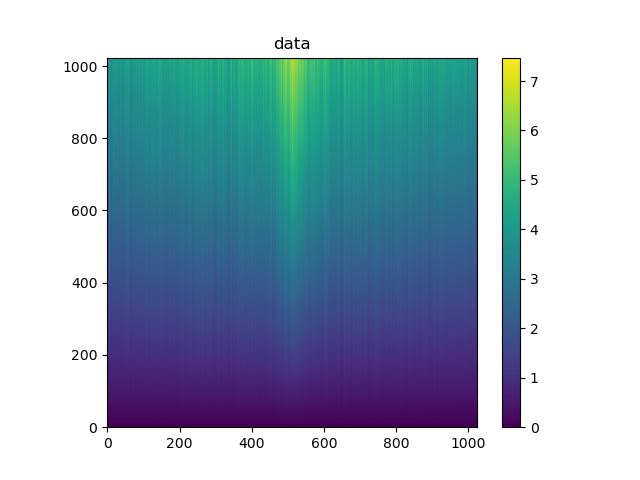}
    \includegraphics[width=0.23\textwidth]{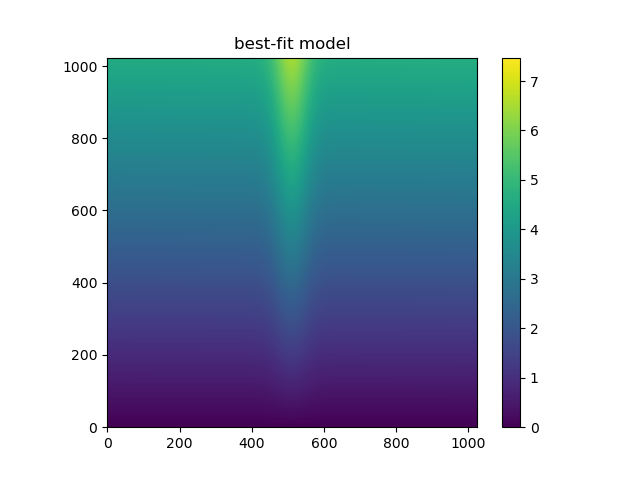}
    \caption{Modeling traps map of an ACIS CCD with a 2-D model defined as ${\rm T}(x,y) = ({\rm a} +  {\rm b}x) y + {\rm c} y * {\rm exp}^{-(x-x0)^2 \over 2 \sigma^2}$. Left panel shows the data and the right panel shows the model image with the color scale on the right in the units of counts.}
    \label{fig:2dmodel}
\end{figure}

\begin{figure}
    \centering    
    \includegraphics[width=0.5\textwidth]{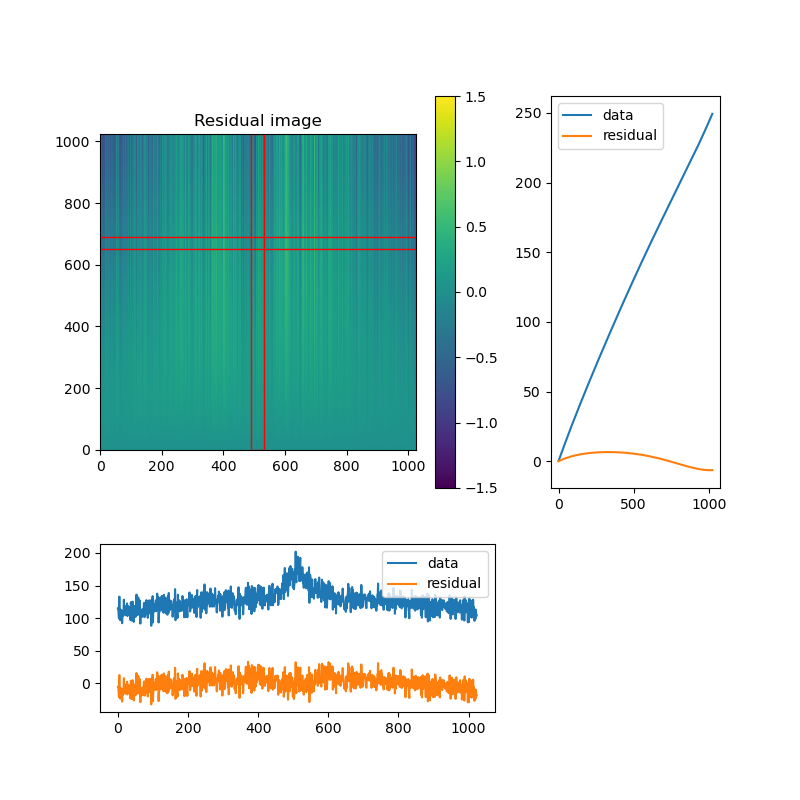}
    \caption{The color image of residuals from the best model shown  in Fig.~\ref{fig:2dmodel}. The color scale indicates $\sigma$ values. The red boxes show the regions used for the 1-d comparison plots shown in the right and bottom panels. The blue lines indicate the data and the orange lines the residuals.}
    \label{fig:2dtrap}
\end{figure}

An example of a user-defined 2-D model function fit to the trap map of a CCD detector is displayed in Figure~\ref{fig:2dmodel} and Figure~\ref{fig:2dtrap}. 
Such models are used to gather information about the characteristics of a detector and can be fit to the data over time for example to monitor changes in a detector performance.

\section{Sherpa and Python ecosystem}
\label{sec:ecosystem}

\begin{figure*}
    \centering
    \includegraphics[width=.9\textwidth]{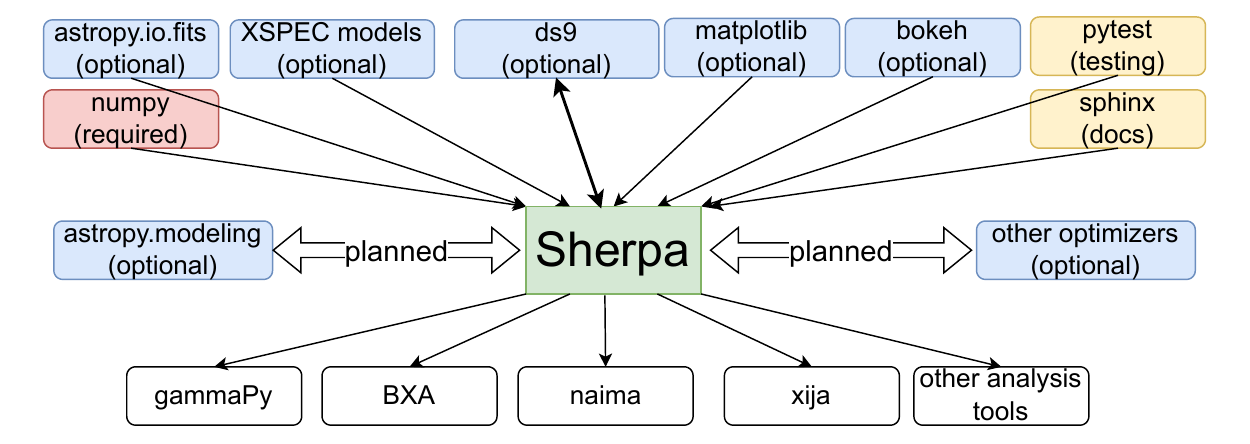}
    \vspace{5pt}
    \caption{\sherpa's place in the scientific Python ecosystem. \sherpa\ requires a numeric library (red box), and interfaces with several optional libraries for I/O, external model libraries and visualization (blue boxes). 
    It enables the fitting and modeling in several domain-specific packages (white boxes). The two large white arrows mark future implementation of interoperability with \texttt{astropy.modeling} and external optimizers. 
    Infrastructure packages needed for developers, but not users, are marked with yellow boxes.}
    \label{fig:ecosystem}
    \vspace{5pt}
\end{figure*}

\sherpa's place in the scientific Python ecosystem is shown in Fig.~\ref{fig:ecosystem}.
\sherpa\ has only a small number of required dependencies, most importantly \texttt{numpy} \citep{numpy}, the foundational numerical array-processing library for all of the scientific Python stack. \sherpa\ also makes use of a number of common Python packages for infrastructure that are used for the installation, for running tests (\texttt{pytest} and some of its plug-ins), and for building the documentation (\texttt{sphinx}). 
As described in Sec.~\ref{sec:visualization} \sherpa\ interfaces with visualization packages \texttt{matplotlib}, \texttt{bokeh}
and \texttt{ds9} and also connects via a plug-in into \texttt{ds9}'s analysis menu \citep{DAX}. 

In turn, \sherpa{} provides functionality for other more specialized packages, that are used for specific astronomical analysis tasks. Those packages may call on \sherpa{} for example to read spectra in specific file formats, to use the \sherpa{} model library and interface to XSPEC models, or to perform numerical model fitting. 
The most well known packages that depend on \sherpa{} are: (1)~\texttt{gammapy} \citep{gammapy:2023}, a Python package for $\gamma$-ray astronomy used as a core-library for the CTA observatory but that also supports other high-energy facilities; (2)~BXA \citep[Bayesian X-ray Analysis, ][]{BXA} a package which connects \sherpa\ (or XSPEC) to a Bayesian nested sampling algorithm; (3) \texttt{naima}, a package to derive non-thermal particle distributions from relativistic electron populations; (4) \texttt{xija}, a package used to model complex time series data using a network of coupled nodes (one particular application of this package is the thermal modeling of the \chandra{} spacecraft).

\section{Project Management}\label{sec:manage}

\subsection{Governance and development model}

The \sherpa\ source code is developed on GitHub to facilitate an open development model. Issues and bugs are tracked and discussed publicly opening the opportunity for contributions. 
Individual code contributions are submitted as pull requests on GitHub, where their utility is reviewed.
The \sherpa\ project requires a minimum of one approval
by a core developer before a pull request is merged into the code base.
\sherpa\ is released twice a year.

\subsection{Collaboration with related projects}
\smallskip

\sherpa\ interfaces with the XSPEC model library \citep{XSPEC} which provides a set of crucial astrophysical models.
\sherpa\ 
is one of the largest users of the library outside of XSPEC.
The \sherpa\ team 
often identifies issues or inconsistencies and provides feedback to XSPEC team.

The \texttt{astropy.modeling} sub-package,
designed like \sherpa, distinguishes data, models, statistics, 
and numerical optimizers. 
However, \sherpa\
has a far larger library of model functions, more advanced optimizers, and more specialized statistics functions. In collaboration with core developers from \texttt{astropy}, the \sherpa\ team developed a package called \texttt{saba}\footnote{\url{https://saba.readthedocs.io/en/latest/}} to make \sherpa\ models, statistics, and optimizers available as plug-ins to the \texttt{astropy} community.

\subsection{Inclusive practices and Sustainability}
\smallskip

Beyond the adoption of standard tools like \texttt{pytest}, \texttt{sphinx}, \texttt{pip} and \texttt{conda}, a large fraction of the code is covered by the test suite. This assures
that a contributed code change does not introduce bugs and break \sherpa.
These efforts make it easier to contribute to \sherpa\ development. 

Code sustainability, as demonstrated by \texttt{astropy} \citep{astropy_II}, requires growing the community of contributors and increasing the size and impact of contributions from outside the core team. 
Currently,
a large fraction of \sherpa\ questions, bug reports, and code contributions are initiated through the CXC helpdesk, with a smaller number of such contributions via GitHub.
\sherpa\ follows GitHub's list of \emph{recommended community standards}\footnote{\url{https://github.com/sherpa/sherpa/community}} to align the project with the best practices for the open development. 
Additionally, developing maintainable and well documented code with
good test coverage, and focus on timely updates to Python and other dependencies
are necessary for long-term sustainability.
Growing the community of \sherpa\ users who become \emph{contributors}, by bringing new ideas and feature requests, will increase sustainability of the project.

\bigskip\bigskip\bigskip

\section{Summary}

\sherpa\ was designed and developed to support traditional forward fitting methodology.
Recent advances in techniques and open source software availability bring new ideas for modeling large complex data sets, including simulation based inference \citep[e.g.,][]{Huppenkothen2022,Barret2024} and machine learning methods \citep[e.g.,][]{Rhea2021, Rhea2023}. These emerging methods may be able to handle multi-dimensional data, such as spatial-spectral-temporal data available in modern observations \citep[see e.g.,][]{Xu2021}. \sherpa\ future development plans include extensions to multi-dimensional data and data cubes from IFUs (Integral Field Units) obtained by modern optical telescopes, such as, for example, James Webb Space Telescope (JWST) and the Giant Magellan Telescope (GMT). 
\sherpa\ flexible framework allows also for incorporating new emerging methods, e.g. simulation based inference for working with complex models. We note that \sherpa's\ open source development model provides a great way for the community contributions to bring new ideas into this modeling framework.

\sherpa\ is a Python package for modeling and fitting astronomical data. Although it has been developed for modeling X-ray data from the beginning, it was designed to model any type of data and it is well suited for multi-wavelength analysis. \sherpa\ has a solid foundation. It has been widely used and tested in many applications over the past two decades. It provides robust methodology to model general astronomical data and its flexibility allows users to  incorporate \sherpa\ in Python scripts or larger data analysis pipelines.

\section{Acknowledgments}

We acknowledge contributions to \sherpa\ by many people in addition to the present authors who have been involved in the project at different stages over the years. The initial design in the late 1990-ties involved Maureen Conroy, Stephen Doe, Malin Ljungberg, and William Joye. Stephen Doe was with the project as the lead system developer until 2013. Peter Freeman focused on developing the scientific modeling and statistical inference methodology (2000-2003), Dan Nguyen (2005-2022) developed optimization methods including implementation of Nelder-Mead and differential evolution algorithms, Chris Stawarz (2005-2006) and Brian Refsdal (2006-2011) worked on the Python version of the code. We also acknowledge discussions and input by current and former members of the Science Data System team  Nina Bonaventura, John Davis, Martin Elvis, Liz Galle, Kenny Glotfelty, John Houck, David P. Huenemoerder and Michael Nowak. Tom Aldcroft, Jamie Budynkiewicz, Johannes Buchner, Christoph Deil, Axel Donath, Iva Laginja, Katrin Leinweber and Brigitta Sipőcz contributed to the open source code via GitHub pull requests. We also thank Janet Evans, Pepi Fabbiano, Margarita Karovska, Vinay Kashyap, Fabrizio Nicastro, Rafael Martinez Galarza and Andreas Zezas for discussions of \sherpa's applications to a variety of data problems.

This work was supported  NASA contract to the Chandra X-ray Center NAS8-03060; HMG and DAP specifically were supported by contract SV3-73016 to MIT for support of the {\em Chandra} X-Ray Center (CXC),
which is operated by SAO for and on behalf of NASA.

This research has made use of NASA’s Astrophysics Data System.

%

\vspace{5mm}
\facilities{\chandra\ X-ray Observatory, NICER, SDSS}


\software{\sherpa, matplotlib}






\bibliography{allrefs}{}
\bibliographystyle{aasjournal}



\end{document}